\documentclass[aps,prc,twocolumn,superscriptaddress,showpacs,floatfix,nofootinbib]{revtex4-1}
\usepackage{amsmath,graphicx}

\begin{document}

\title{Anisotropic flow in event-by-event ideal hydrodynamic 
simulations of $\sqrt{s_{NN}} $=200~GeV Au+Au collisions}

\author{Fernando G.  Gardim}
\author{Fr\'ed\'erique Grassi}
\affiliation{
Instituto de F\'\i sica, Universidade de S\~ao Paulo, C.P. 66318, 05315-970, S\~ao Paulo-SP, Brazil}
\author{Matthew Luzum}
\author{Jean-Yves Ollitrault}
\affiliation{
CNRS, URA2306, IPhT, Institut de physique th\'eorique de Saclay, F-91191
Gif-sur-Yvette, France}
\date{\today}

\begin{abstract}
We simulate top-energy Au+Au collisions using ideal hydrodynamics
in order to make the first  comparison to the complete set of mid-rapidity flow measurements made by the PHENIX Collaboration.  
A simultaneous calculation of $v_2$, $v_3$, $v_4$, and the first event-by-event calculation of quadrangular flow defined with respect to the $v_2$ event plane ($v_4\{\Psi_2\}$) gives 
good agreement with measured values,
including the dependence on both transverse momentum and centrality.    
This provides confirmation that the collision system is indeed well described as a 
quark-gluon plasma with an extremely small viscosity, and that correlations 
are dominantly generated from collective effects.
In addition we present a prediction for $v_5$.
\end{abstract}


\maketitle

\section{Introduction}
Evidence suggests that in a collision between ultra-relativistic heavy nuclei, a strongly-interacting, low-viscosity fluid  --- the quark-gluon plasma (QGP) --- is created.  The clearest indication of this behavior is seen in the azimuthal anisotropy~\cite{Ollitrault:1992bk} among the bulk of emitted particles.  In theory, one characterizes this anisotropy in terms of a single-particle probability distribution for each collision event.  By writing this distribution as a Fourier series with respect to the azimuthal angle of outgoing particles $\phi$, one can define flow coefficients $v_n$ and event plane angles $\Psi_n$:
\begin{align}
\label{eq1}
\frac{2\pi}{N} \frac {dN} {d\phi} &= 1 + 2 \sum_{n=1}^\infty v_n \cos n(\phi - \Psi_n) , \\
v_n e^{i n \Psi_n} &\equiv \langle e^{i n \phi} \label{defvn}\rangle ,
\end{align}
where the brackets indicate an average over the single particle
probability, and the event plane angles $\Psi_n$ are chosen such that $v_n$ are the (positive) magnitudes of the complex Fourier coefficients.  

Experimentally, one measures the azimuthal dependence of event-averaged correlations between detected particles.  These measurements indicate the presence of a very large ``elliptic flow'' coefficient $v_2$~\cite{Ackermann:2000tr,Aamodt:2010pa}, which typically can only be reproduced in calculations where the system is modeled as a strongly-interacting fluid.  In this picture, the large momentum anisotropy is generated as a hydrodynamic response to the spatial anisotropy of the nuclear overlap region in collisions of non-zero impact parameter.  It even appears that the created quark-gluon plasma must be an almost perfect (zero viscosity) fluid, with a ratio of shear viscosity to entropy density $\eta/s$ that is at most a few times $1/(4\pi)$, a value that was famously conjectured to be a universal lower bound~\cite{Kovtun:2004de}\footnote{The bound is now known to be violated in some theories~\cite{Brigante:2007nu,Kats:2007mq,Buchel:2008vz}, and it may even be possible to have an arbitrarily small value~\cite{deBoer:2009gx}, though the \textit{effective} viscosity may still have a finite bound~\cite{Kovtun:2011np}.}.
However, the extraction of a precise finite value is hampered by poor knowledge of the earliest stages of the collision, as well as other uncertainties~\cite{Luzum:2008cw}.

An important recent development was the realization of the importance
of quantum fluctuations, which in particular implies an event-by-event breaking of the symmetry naively implied by the collision of identical nuclei.  Specifically, the
coefficients $v_n$ are generally non-zero also for odd
$n$~\cite{Alver:2010gr},  the event plane angles do not necessarily point in the same direction as the impact parameter~\cite{Alver:2006wh, Andrade:2006yh},  and these quantities fluctuate significantly from one event to another, even at a fixed impact parameter~\cite{Miller:2003kd}.

These insights led to the possibility that \textit{all} of the measured long-range correlations may be generated solely from collective behavior~\cite{Alver:2010gr,Luzum:2010sp}.

Several new flow observables --- specifically ones implied by the presence of event-by-event fluctuations --- were recently measured for the first time~\cite{Adare:2011tg,ALICE:2011ab,collaboration:2011hfa,Sorensen:2011fb,Chatrchyan:2012wg}.  Studies of these new observables indicate that, individually, they indeed appear to have properties that are consistent with a hydrodynamic origin~\cite{Alver:2010dn,Schenke:2010nt,Luzum:2010sp,Sorensen:2011hm}.  However, they have not yet all been reproduced in a single calculation within one model using a single set of parameters.  This has left some lingering doubt about whether the interpretation in terms of collective behavior is indeed correct~\cite{INT}.  In addition, each measurement provides an independent constraint on theory, so identifying models and sets of parameters that can simultaneously satisfy all the constraints is a necessary first step in reducing various theoretical uncertainties.

In this Letter we perform state-of-the-art ideal hydrodynamic
calculations and compare the results to the first
measurements~\cite{Adare:2011tg} of these new observables at
Relativistic Heavy-Ion Collider (RHIC) as well as previous
measurements by the same collaboration~\cite{Adare:2010ux}. 
Other groups have presented calculations from some of these
observables using event-by-event
ideal~\cite{Werner:2010aa,Holopainen:2010gz,Qiu:2011iv,Petersen:2012qc} or
viscous~\cite{Schenke:2011bn} hydrodynamics, or transport
models~\cite{Konchakovski:2012yg}. 
The present study encompasses simultaneously, for the first time, all 
the measured flow observables at midrapidity.

\section{Observables}
\label{s:observables}

All the experimental results considered here were obtained
using the event-plane
method~\cite{Poskanzer:1998yz}.  With this method, one first
identifies an event plane $\Psi_n$ in each event using a specific
detector at forward rapidity, and then calculates the correlation of
particles near midrapidity with this event plane, e.g.,  
\begin{equation}
\label{defEP}
v_n\{\Psi_n\}\equiv \langle\cos n(\phi - \Psi_n)\rangle,  
\end{equation}
where the brackets indicate an average over particles in a large number
 of events. 
 A rapidity gap with the event-plane detector suppresses
nonflow correlations~\cite{Luzum:2010sp,Aamodt:2011by}. 
At RHIC, ``triangular flow'' $v_3\{\Psi_3\}$ and ``quadrangular flow'' $v_4\{\Psi_4\}$ were measured for the first time, as a function of the particle 
transverse momentum $p_t$ in various centrality classes by
the PHENIX collaboration~\cite{Adare:2011tg} (preliminary data from STAR have now also been 
presented~\cite{Sorensen:2011fb}).  

Previously, a different quadrangular flow observable has been measured, defined with respect to
$\Psi_2$~\cite{Adare:2010ux, Adams:2003zg}. We use a different notation for this
quantity to avoid confusion: 
\begin{equation}
\label{defv42}
v_4\{\Psi_2\}\equiv \langle\cos 4(\phi - \Psi_2)\rangle. 
\end{equation}

 %
%
%
\begin{figure*}
 \includegraphics[width=\linewidth]{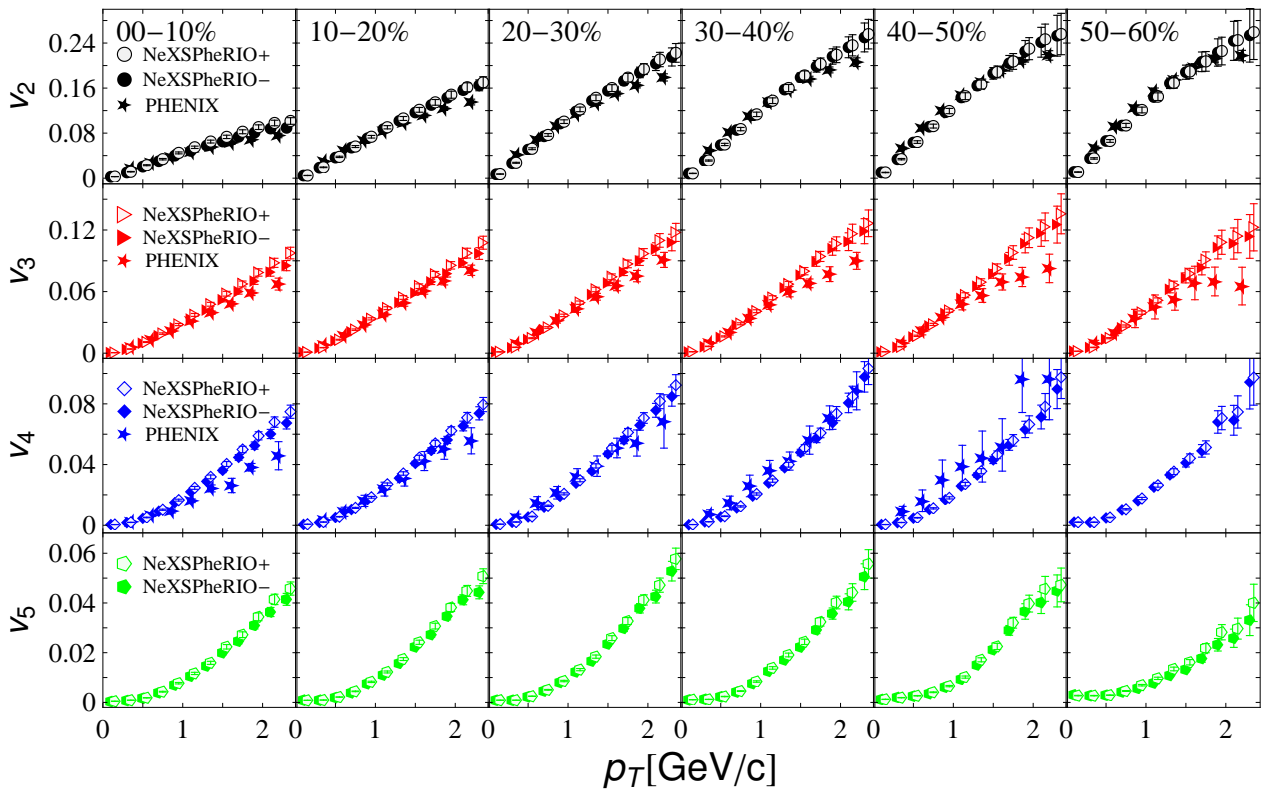}
 \caption{
(Color online) Results for $v_n\{\Psi_n\}$ for $n=$ 2--5, compared to
published data from the PHENIX collaboration~\cite{Adare:2011tg}. 
Closed and open symbols correspond to two different ways of averaging
over events (mean and rms value, respectively).
Error bars represent statistical uncertainty from the finite number
of events. The left column (0--10\%) represents the 10\% most central collisions, which each column to the right increasingly peripheral.}  
\label{fig:vn}
\end{figure*}

$v_n$ is analyzed using a large sample of events,
and its value fluctuates from one event to the other. 
These fluctuations (which were not appreciated when the method was developed), 
combined with the use of a finite
sample of particles in the analysis, cause the measured
value to deviate from the event average of the theoretical 
coefficients defined in Eq.~\eqref{eq1}.
Generally, $v_n\{\Psi_n\}$ lies between the mean value and the
root-mean-square (rms) value of $v_n$.
One can parameterize the resulting
measurement as~\cite{Alver:2008zza}: 
\begin{equation}
\label{alpha}
v_n\{\Psi_n\}\simeq\langle v_n^\alpha \rangle^{1/\alpha}, 
\end{equation}
where here the brackets indicate an average over events. The value of
$\alpha$ depends on the event plane
resolution ${\rm{Res}}\{\Psi_n\}\sim v_n\sqrt{N}$~\cite{Ollitrault:2009ie}: If the resolution is poor,   
$\alpha\simeq 2$, and the measured $v_n$ is a rms value, while if the
resolution is large, $\alpha\simeq 1$, and the result gets closer to
the mean value. 

The most recent data from PHENIX has a maximum event plane resolution
of 0.74 (for $v_2$ around 30\% centrality~\cite{Adare:2010ux}) and
much smaller for $v_3$ and $v_4$~\cite{Adare:2011tg}, which implies
$\alpha>1.81$~\cite{Ollitrault:2009ie}.  So in general the results are
very close to a rms value of $v_n$.  Nevertheless, in the following we compute both limiting cases $\alpha=2$ and $\alpha=1$ in order to show the size of
the effect of fluctuations on event-plane analyses.

Likewise, the measured value $v_4\{\Psi_2\}$ depends on the
resolution~\cite{Gombeaud:2009ye}, and is usually close to $\langle
v_4 v_2^2 \cos (4\Psi_4-4\Psi_2) \rangle/\sqrt{ \langle v_2^4 \rangle}$, but
with increasing resolution approaches $\langle v_4
\cos(4\Psi_4-4\Psi_2)\rangle$. 
\section{Results}
\label{s:results}
\begin{figure*}
 \includegraphics[width=\linewidth]{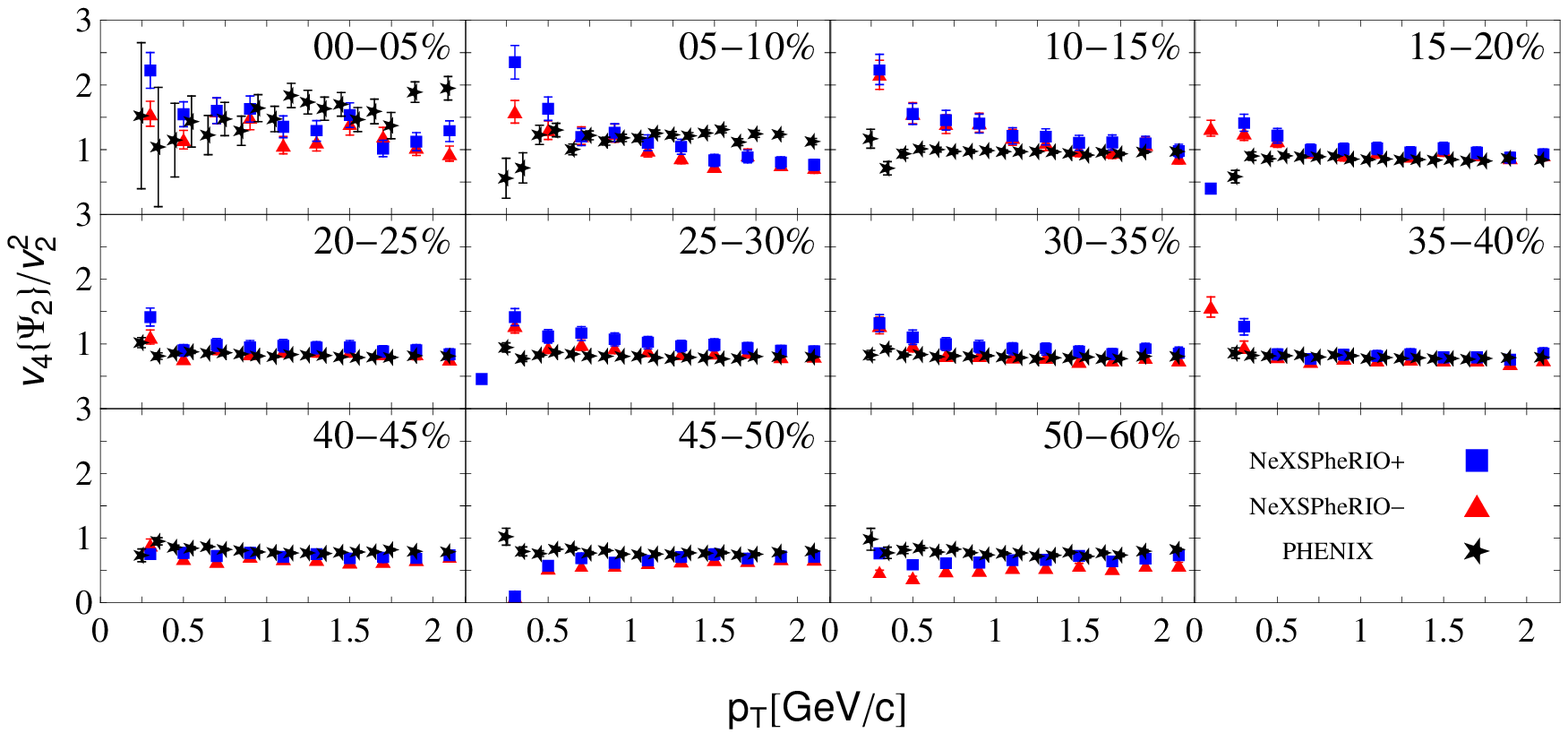}
 \caption{
(Color online)  Results for the first event-by-event hydrodynamic calculation of 
$v_4\{\Psi_2\}/v_2\{\Psi_2\}^2$, compared to
published data from the PHENIX collaboration~\cite{Adare:2010ux}. 
As in Fig.~\ref{fig:vn}, closed and open symbols correspond to two
different ways of averaging over events (see text), 
error bars represent statistical uncertainty from the finite number
of events, and smaller percentile refers to more central collisions }
\label{fig:v4}
\end{figure*}
Using the hydrodynamic code NeXSPheRIO~\cite{Hama:2004rr}, we simulate
top-energy Au-Au collisions at RHIC.  This code solves the equations
of ideal relativistic hydrodynamics using fluctuating initial
conditions from the event generator NeXus~\cite{Drescher:2000ha}.  

NeXus aims at a realistic and consistent approach of the initial
stage of nuclear collisions~\cite{Drescher:2000ha}. 
It is a Monte-Carlo generator which takes into account not only the
fluctuations of nucleon positions within nuclei~\cite{Schenke:2011bn},
but also fluctuations at the partonic level: the momentum of each
nucleon is shared between  one or several ``participants'' and a
``remnant'', which implies non-trivial dynamical fluctuations in each
nucleon-nucleon collisions.  The resulting full energy-momentum tensor 
is matched to a hydrodynamic form, resulting in a fluctuating flow field 
in addition to a fluctuating initial energy density, in all three spatial dimensions,
with the transverse length scale of the fluctuations set mostly by the size of the 
incoming nucleons.

At the end of the hydrodynamic evolution, discrete particles are
emitted using a Monte-Carlo 
generator\footnote{Freeze-out occurs at a constant
  temperature. Hadrons do not rescatter after 
  freeze-out~\cite{Teaney:2001av,Petersen:2008dd,Song:2011hk}, but resonance decays are implemented.}.
NeXSPheRIO provides a good description of rapidity and transverse
momentum spectra~\cite{Qian:2007ff}, elliptic flow $v_2$~\cite{Andrade:2008xh}, and the rapidity-even $v_1$ observable, directed flow at midrapidity~\cite{Gardim:2011qn}.
In addition, it is known to reproduce the  long-range structures observed in two-particle
correlations~\cite{Takahashi:2009na}.  All parameters were fixed from these earlier investigations, 
before any of the new observables ($v_3$, $v_4$) were measured --- nothing has been tuned here.

For this work, we generated 110 NeXus events each in 5\%  centrality classes up to 60\% centrality,
 solving the hydrodynamic equations independently for each event.   As
 in Ref.~\cite{Gardim:2011xv}, at the end of each hydro event, we run
 the Monte-Carlo generator many times, so that we can do the flow
 analysis using approximately $6\times 10^5$ particles per event.
 This significantly reduces statistical noise and allows for an
 accurate determination of $v_n$ and $\Psi_n$ in every event.  It also suppresses 
 non-flow correlations from, e.g., particle decays.  These quantities are then
 calculated by Eq.~(\ref{defvn}), with the average taken
 over all  particles in the pseudorapidity interval $-1<\eta<1$.  
The procedure used to measure $v_n$ in hydrodynamics thus mimics the
experimental procedure, with two differences: (i) there is no need for
a rapidity gap, because nonflow correlations are negligible; (ii) there is no need for a resolution correction,
because the huge multiplicity per event ensures that the resolution is
close to 1 for all events~\cite{Gardim:2011xv}.  

Fig.~\ref{fig:vn} displays $v_n$ as a function of the particle
transverse momentum $p_t$, averaged over events in a centrality
class. 
The average over events is estimated in two different ways in order to
illustrate the effect of event-by-event flow fluctuations on the experimental analysis. 
The first estimate, labeled NeXSPheRIO-, is a plain mean value
(corresponding to $\alpha=1$ in Eq.~\eqref{alpha}).
The second estimate, labeled NeXSPheRIO+ is
a weigthed average
\begin{equation}
\label{defplus}
v_n^+\{\Psi_n\}\equiv \frac{\langle v_n\cos n(\phi -
  \Psi_n)\rangle}{\sqrt{\langle v_n^2\rangle}}.  
\end{equation}
The average of $v_n^+\{\Psi_n\}$ over $p_t$ is the rms average of
$v_n$ ($\alpha=2$ in Eq.~\eqref{alpha}). 
For Gaussian flow fluctuations~\cite{Voloshin:2007pc}, 
the ratio of the rms to the mean is $\sqrt{4/\pi}\simeq 1.13$ 
for $v_3$ and $v_5$, and closer to 1 for $v_2$ and $v_4$. 

Fig.~\ref{fig:vn} shows that our event-by-event ideal
hydrodynamic calculation reproduces well the observed centrality and transverse
momentum dependence of $v_2$, $v_3$ and $v_4$. 
The $p_t$ dependence is a generic feature of ideal hydrodynamics~\cite{Alver:2010dn}. 
The magnitude and centrality dependence of $v_n$, on the other hand,
depend on the initial conditions: $v_2$ is mostly
driven by the almond shape of the overlap area, which depends on the
particular model used~\cite{Lappi:2006xc},
while higher harmonics are mostly driven by initial
fluctuations~\cite{Alver:2010gr}, which explains why they have a mild
centrality dependence~\cite{Bhalerao:2011bp}.
%

Originally, quadrangular flow $v_4$ had been measured with respect to the 
event-plane of elliptic flow. Recent results show that 
$v_4\{\Psi_2\}$~\cite{Adare:2010ux} is smaller than 
$v_4\{\Psi_4\}$~\cite{Adare:2011tg}, typically by a factor 2 for
peripheral collisions, and by a factor 5 for central collisions. 
Although $v_4\{\Psi_2\}$ is smaller, it is measured with a
better relative accuracy than $v_4\{\Psi_4\}$, because of the better
resolution on $\Psi_2$. This makes $v_4\{\Psi_2\}$ a useful quantity
for detailed model comparisons~\cite{Luzum:2010ae}. 
As in the case of $v_n\{\Psi_n\}$, we perform the average over events
in two different ways in order to illustrate the effect of
event-by-event flow fluctuations.  
The first estimate, labeled NeXSPheRIO-, is a plain mean value, as in
Eq.~(\ref{defv42}). The second estimate, labeled NeXSPheRIO+ is
a weigthed average:
\begin{equation}
v_4^+\{\Psi_2\}\equiv
 \frac{\langle v_2^2\cos 4(\phi -
  \Psi_2)\rangle}{\sqrt{\langle v_2^4\rangle}}.  
\end{equation}
The actual event-plane value is expected to lie between
these two limits, depending on the resolution~\cite{Gombeaud:2009ye}. 

Since $v_4\{\Psi_2\}$ can be generated by elliptic flow as a 
second order effect~\cite{Borghini:2005kd}, we scale it by
$v_2\{\Psi_2\}^2$ for each $p_t$. Hereafter, we denote 
 $v_4\{\Psi_2\}$ and $v_2\{\Psi_2\}$ simply by $v_4$ and $v_2$. 
Fig.~\ref{fig:v4} displays this first event-by-event hydrodynamic calculation 
of $v_4/v_2^2$ as a
function of $p_t$ for different centralities. 
The measured ratio is remarkably constant as a function of
$p_t$, and increases mildly for central collisions. 
Ideal hydrodynamics predicts 
$v_4/v_2^2\simeq1/2$ at high $p_t$ for a single
event~\cite{Borghini:2005kd}. 
For all centralities, the measured value of $v_4/v_2^2$ is greater
than $1/2$, even at high $p_t$. 
This can be explained~\cite{Gombeaud:2009ye}  
by $v_2$ fluctuations, except for the two most central bins,
where one expects $v_4/v_2^2\simeq1$~\cite{Gombeaud:2009ye}, smaller than
the measured value, which is between $1.5$ and $2$ for the most
central bin. 
For these two central bins, our results from event-by-event
hydrodynamics are in good agreement with experiment (first two panels
in Fig.~\ref{fig:v4}). 
This shows that other sources of flow fluctuations, other than $v_2$
fluctuations alone, contribute to $v_4/v_2^2$. 
A similar finding has been reported in a transport
calculation with $v_4$ and $v_2$ both defined with respect to the direction of
the impact parameter~\cite{Konchakovski:2012yg}. 
Our calculated $v_4/v_2^2$ is slightly higher than data for the next
two bins ($10-20\%$). Above $20\%$ centrality, data are within the
range spanned by our calculations. 

The calculations shown here simulate the system evolution using ideal hydrodynamics, i.e., with negligible viscosity.
These results prove that no non-zero QGP viscosity is required to reproduce these data.  In fact, this calculation \textit{requires} a negligible viscosity --- keeping everything else fixed, a viscosity the size of the conjectured bound $\eta/s=1/4\pi$ would suppress $v_n$ and destroy the remarkable fit to data.    In addition, the ratio $v_4/v_2^2 (p_T)$ depends strongly on $\eta/s$, and any non-zero value usually tends to destroy the flat curve that ideal hydrodynamics predicts~\cite{Luzum:2010ad,Luzum:2010ae}.    

However,  this requirement of negligible viscosity depends crucially on aspects of the model which are not entirely constrained.
In particular, although the NeXus model provides an honest effort at a reasonable description of the physics, with many realistic elements, there is considerable uncertainty about the early stages of a heavy-ion collision and the resulting initial conditions for hydrodynamic evolution.  In principle, another model, coupled to viscous hydrodynamics might well be able to fit these data.  For example Ref.~\cite{Schenke:2011bn} presents event-by-event viscous hydrodynamic calculations with Glauber initial conditions that require a value close to $\eta/s=0.08$ to give reasonable agreement with the quantities in Fig.~\ref{fig:vn} at several centralities, though they underpredict $v_3$ for central collisions.  Secondly, although $v_4/v_2^2$ is not very sensitive to the initial conditions, the effect of non-zero viscosity depends significantly on the way it is implemented at freeze out~\cite{Luzum:2010ad},
and the correct implementation is an open issue.
 
Thus, this work is only a first step in identifying models that are compatible with data, and strong conclusions cannot yet be drawn about, e.g., the precise value of $\eta/s$.  Although the success of these calculations are an important milestone, proving that at this point no lower bound can yet be placed on $\eta/s$, we can not yet make a precise statement about an upper bound -- only that it still appears unlikely that a value significantly larger than $1/4\pi$ will be possible.

\section{Conclusions}
Using an ideal hydrodynamic model with
fluctuating initial conditions, we have performed
the first simultaneous calculation of 
$v_2\{\Psi_2\}$, $v_3\{\Psi_3\}$, $v_4\{\Psi_4\}$ and $v_4\{\Psi_2\}$
as a function of transverse momentum and centrality.
%
Our results are in good agreement with the most recent experimental results
for all the observables at RHIC,  
at all centralities and in a wide range of transverse momentum.
This provides convincing confirmation of the current paradigm that collective effects alone can explain
all long-range correlations in the soft sector.  Further, since all such
measured correlations are generated consistently in a single calculation,  this provides 
a complete, unified picture of the bulk evolution of a heavy-ion collision 
as an extremely low-viscosity fluid.
Indeed, for our model of initial conditions,
a negligible viscosity is required for a good fit to all mid-rapidity flow observables.
Therefore no lower bound can currently be placed on the shear viscosity of the quark-gluon plasma.
Further study will be needed to determine a reliable upper bound, but finding 
models (such as this one) that are compatible with all measured data is a significant first step.

\begin{acknowledgments}
We thank Yogiro Hama for useful discussion.  This work is funded by 
Cofecub under project Uc Ph 113/08;2007.1.875.43.9, by FAPESP under projects 09/50180-0 and 09/16860-3, and by CNPq under project  301141/2010-0. ML is supported by the European Research Council under the
Advanced Investigator Grant ERC-AD-267258.
\end{acknowledgments}

\end{document}